\input{aipcheck}

\documentclass[
    ,final            
  ]
  {aipproc}
\usepackage{amsmath,amsfonts,amssymb}
\usepackage{eucal}
\def\openone{\leavevmode\hbox{\small1\kern-3.3pt\normalsize1}}

\def\bbbz{{\Bbb Z}}
\def\openone{\leavevmode\hbox{\small1\kern-3.3pt\normalsize1}}
\def\bbbz{{\Bbb Z}}

\layoutstyle{6x9}

\def\ad{\mbox{ad\,}}

\def\diag{\mbox{diag\,}}
\def\fr#1{{\mathfrak{#1}}}

\def\openone{\leavevmode\hbox{\small1\kern-3.3pt\normalsize1}}

\def\m{{\mathbf m}}

\def\ad{\mathrm{ad\,}}

\def\bdiag{\mbox{block-diag\,}}
\def\diag{\mbox{diag\,}}

\def\bbbe\mathbb{E}

\def\bbbc{\mathbb{C}}
\def\bbbe{\mathbb{E}}
\def\bbbz{\mathbb{Z}}

\begin{document}

\title{On Reductions of Soliton Solutions of multi-component NLS models
and Spinor Bose-Einstein condensates }


\author{V. S. Gerdjikov}{
  address={Institute for Nuclear Research and Nuclear
Energy,  Bulgarian Academy of Sciences, 72 Tsarigradsko chaussee
1784 Sofia, Bulgaria}
}



\begin{abstract}
We consider a class of multicomponent nonlinear Schr\"odinger
equations (MNLS) related to the symmetric {\bf BD.I}-type symmetric spaces.
As important particular case of these MNLS we obtain the Kulish-Sklyanin model.
Some new reductions and their effects on the soliton solutions are obtained by
proper modifying the Zakahrov-Shabat dressing method.

\end{abstract}

\maketitle


\section{Introduction}

Consider Bose-Einstein condensate (BEC) of alkali atoms in the $F=1$ hyperfine state,
elongated in $x$ direction and confined in the transverse
directions $y,z$ by purely optical means.
The dynamics of this assembly of atoms is described by a  3-component
normalized spinor wave vector  $ {\vec{\Phi}}=(\Phi_1, \Phi_0 , \Phi_{-1})^{T}(x,t)$ satisfying the
multicomponent nonlinear Schr\"{o}dinger (MNLS) equation \cite{IMW04,86,84}, which in dimensionless
coordinates can be written down as:
\begin{eqnarray}\label{eq:1}
\begin{split}
&& i\partial_{t} \Phi_{\pm 1}+\partial^{2}_{x} \Phi_{\pm 1}+2(|\Phi_{\pm 1}|^2
+ |\Phi_{0}|^2) \Phi_{\pm 1} + \Phi_{\mp 1}^{*}\Phi_{0}^2=0, \\
&& i\partial_{t} \Phi_{0}+\partial^{2}_{x} \Phi_{0}+(2|\Phi_{+1}|^2
+ |\Phi_{0}|^2+ 2|\Phi_{-1}|^2) \Phi_{0} +2\Phi_{0}^{*}\Phi_{+1}\Phi_{-1}=0,
\end{split}
\end{eqnarray}
The second model which describes BEC with $F=2$ hyperfine structure is a 5-component MNLS system:
\begin{eqnarray}\label{eq:2}
\begin{split}
&& i\partial_{t} \Phi_{\pm 2}+\partial^{2}_{x} \Phi_{\pm 2}+2\left( (\vec{\Phi}{\,}^\dag,\vec{\Phi})
 - |\Phi_{\pm 2}|^2 \right)\Phi_{\pm 2}
+ 2 \Phi_{\mp 2}^*\Phi_{+ 1}\Phi_{- 1} - \Phi_{\mp 2}^{*} \Phi_{0}^2=0, \\
&& i\partial_{t} \Phi_{\pm 1}+\partial^{2}_{x} \Phi_{\pm 1}+2\left(
(\vec{\Phi}{\,}^\dag,\vec{\Phi}) - |\Phi_{\pm 1}|^2 \right)\Phi_{\pm 1} + 2\Phi_{\mp 1}^{*}\Phi_{-1} \Phi_{+1}
+ \Phi_{\mp 1}^{*}\Phi_{0}^2=0, \\
&& i\partial_{t} \Phi_{0}+\partial^{2}_{x} \Phi_{0}+2\left(
(\vec{\Phi}{\,}^\dag,\vec{\Phi}) - \frac{1}{2}|\Phi_{0}|^2 \right) \Phi_{0}
 +2\Phi_{0}^{*}\Phi_{+1}\Phi_{-1} - 2\Phi_{0}^{*}\Phi_{+2}\Phi_{-2}=0,
\end{split}
\end{eqnarray}
where $(\vec{\Phi}{\,}^\dag,\vec{\Phi})=\sum_{k=-2}^2 |\Phi_{k}|^2 $.
Both models allow Lax representations and therefore are integrable by the inverse scattering transform
method \cite{ForKu*83,86,84}. The Lax pairs have natural Lie algebraic structure which relates them
to the symmetric spaces ${\bf BD.I}\simeq {\rm SO(n+2)}/{\rm SO(n)\times SO(2)}$ with $n=3$ and $n=5$ respectively.
From algebraic point of view this means that the potential $Q(x,t)$ of $L$ takes the form $Q(x,t) = [J,X(x,t)]$
where $X(x,t)$ is a generic element of the Lie algebra $so(n+2)$ and the constant element $J =
\diag (1,0,\dots , 0 -1)$ is a specially chosen element of the Cartan subalgebra $\fr{h}\subset \fr{g}$. For more
details see \cite{Helg,ForKu*83}.

The present paper extends the results of \cite{86,84} for the class of MNLS related to {\bf BD.I}-type symmetric spaces,
i.e. for any $n$. We briefly outline how the direct and inverse scattering problem for the Lax operator
are reduced to a Riemann-Hilbert problem. Next we find that a simple change of variables can cast the above-mentioned
MNLS into the Kulish-Sklyanin model (KSM) \cite{KuSkl}.
We also apply Mikhailov reduction group method \cite{Mikh} and derive several new types of MNLS interactions.
We derive also the constraints on the polarization vectors in the dressing factors that are imposed by the reductions.
Finally we apply a proper modification (see  \cite{86,84}) of the Zakharov-Shabat dressing method \cite{ZaSh,ZaMi} and
derive the soliton solutions of the MNLS and of KSM in particular.   Thus we obtain several
new types of integrable vector MNLS and their soliton solutions.

The majority of papers devoted to soliton equations analyze and solve the inverse scattering problem (ISP)
for the relevant Lax operators using the typical (lowest dimensional) representation
of the corresponding Lie algebra. At the end of our paper we briefly compare the  properties of
the dressing factors in two of the fundamental representations of the Lie algebra $so(2r)$. We
also elucidate some additional issues considered in \cite{84,82} such as the structure of the soliton solutions
and the effect of additional $\bbbz_2$-reductions.

\section{MNLS  equations for {\bf BD.I} series of symmetric spaces}

MNLS equations  for the {\bf BD.I.} series of symmetric spaces
(algebras of the type $so(n+2)$ and $J$ dual to $e_1$) have the
Lax representation $[L,M]=0$ as follows
\begin{eqnarray}\label{eq:3.1}
L\psi (x,t,\lambda ) &\equiv & i\partial_x\psi + (Q(x,t) - \lambda
J)\psi  (x,t,\lambda )=0.\\ \label{eq:3.2} M\psi (x,t,\lambda )
&\equiv & i\partial_t\psi + (V_0(x,t) +
\lambda V_1(x,t) - \lambda ^2 J)\psi  (x,t,\lambda )=0, \\
V_1(x,t)&=& Q(x,t), \qquad V_0(x,t) = i \ad_J^{-1} \frac{d Q}{dx}
+ \frac{1}{2} \left[\ad_J^{-1} Q, Q(x,t) \right].
\end{eqnarray}
where $\ad_J X=[J,X]$ and $\ad_J^{-1}$ is well defined  on the image of $\ad_J$ in $\fr{g}$;
\begin{equation}\label{vec1}
Q=\left(\begin{array}{ccc}  0 & \vec{q}^{T} & 0 \\
  \vec{p} & 0 & s_{0}\vec{q} \\  0 & \vec{p}^{T}s_{0} & 0 \\
\end{array}\right),\qquad J=\mbox{diag}(1,0,\ldots 0, -1).
\end{equation}
The $n$-component vectors $\vec{q}$ and $\vec{p}$ have the form
\begin{equation}\label{eq:sok}
\vec{q} = (q_1,\dots ,  q_{n})^T, \qquad \vec{p} = (p_1,\dots , p_{n})^T,
\end{equation}
while the matrix $s_0 =S^{(n)}$ enters in the definition of
$so(n)$:
\begin{equation}\label{eq:z1.6a}
 X\in so(n), \qquad X + S^{(n)} X^T S^{(n)} =0, \qquad S^{(n)} =
 \sum_{s=1}^{n} (-1)^{s+1} E_{s, n+1-s}^{(n)},
\end{equation}
for $n=2r+1$ and
\begin{equation}\label{eq:z1.6b}
S^{(n )} = \sum_{s=1}^{r} (-1)^{s+1} (E_{s, n+1-s}^{(n)}+E_{n+1-s,s}^{(n)})
\end{equation}
for $n=2r$. By $E^{(n)}_{sp}$ above we mean $n\times n$ matrix whose matrix
elements are $(E^{(n)}_{sp})_{ij}=\delta_{si}\delta_{pj}$.
With the definition of orthogonality used in (\ref{eq:z1.6a}) the Cartan generators $H_k=E^{(n)}_{k,k}
-E^{(n)}_{n+1-k,n+1-k}$ are represented by diagonal matrices.

The Lax pairs, related to the symmetric spaces $SO(n+2)/(SO(n)\times
SO(2))$ have special algebraic properties. They are determined by choosing $J=H_1$ to be dual
to $e_1 \in \bbbe^r$. It allows one to introduce a grading in  $\fr{g}$, i.e.
$\mathfrak{g}=\mathfrak{g}_0\oplus \mathfrak{g}_1$ so that:
\begin{equation}\label{eq:grad}
[X_1, X_2]\in \mathfrak{g}_0, \qquad [X_1,Y_1]\in \mathfrak{g}_1,
\qquad [Y_1,Y_2]\in \mathfrak{g}_0,
\end{equation}
for any choice of the elements $X_1, X_2\in \mathfrak{g}_0$ and
$Y_1, Y_2 \in \mathfrak{g}_1$. The grading splits the set of
positive roots of $so(n)$ into two subsets $\Delta^+= \Delta_0^+
\cup \Delta_1^+$ where $\Delta_0^+$ contains all the positive
roots of $\mathfrak{g}$ which are orthogonal to $e_1$, i.e.
$(\alpha,e_1)=0$; the roots in $\beta\in\Delta_1^+$ satisfy
$(\beta,e_1)=1$. For more details see \cite{Helg}.

In writing down the Lax pair (\ref{eq:3.1}) we made use of the
typical $n\times n$ representation of $so(n)$. The Lax pair can be
considered in any representation of $so(n)$, then the potential
$Q$ will take the form:
\begin{equation}\label{eq:QQ}
Q(x,t) =\sum_{\alpha\in\Delta_1^+} \left( q_{\alpha}(x,t)
E_{\alpha} + p_{\alpha}(x,t) E_{-\alpha} \right).
\end{equation}

Next we introduce $n$-component `vectors' formed by the Weyl
generators of $so(n+2)$ corresponding to the roots in
$\Delta_1^+$:
\begin{equation}\label{eq:Epm}
\vec{E}_1^\pm = ( E_{\pm (e_1-e_2)}, \dots , E_{\pm (e_1-e_r)},
E_{\pm e_1}, E_{\pm(e_1+e_r)}, \dots , E_{\pm (e_1+e_2)} ),
\end{equation}
for $n=2r+1$ and
\begin{equation}\label{eq:Epmb}
\vec{E}_1^\pm = ( E_{\pm (e_1-e_2)}, \dots , E_{\pm (e_1-e_r)},
E_{\pm(e_1+e_r)}, \dots , E_{\pm (e_1+e_2)} ),
\end{equation}
for $n=2r$. Then the generic form of the potentials $Q(x,t)$
related to these type of symmetric  spaces can be written as sum
of two "scalar" products
\begin{equation}\label{eq:Q}
Q(x,t) = (\vec{q}(x,t) \cdot \vec{E}_1^+) + (\vec{p}(x,t) \cdot
\vec{E}_1^-) .
\end{equation}

In terms of these notations the generic MNLS type equations
connected to ${\bf BD.I.}$ acquire the form
\begin{equation}\label{eq:4.2}
\begin{split}
i \vec{q}_t &+ \vec{q}_{xx} + 2 (\vec{q},\vec{p}) \vec{q} -
(\vec{q},s_0\vec{q}) s_0\vec{p} =0, \\
i \vec{p}_t &- \vec{p}_{xx} - 2 (\vec{q},\vec{p}) \vec{p} +
(\vec{p},s_0\vec{p}) s_0\vec{q} =0,
\end{split}
\end{equation}
With the typical reduction $p_k=q_k^{*}$ it gives:
\begin{equation}\label{eq:4.2'}
i \vec{q}_t + \vec{q}_{xx} + 2 (\vec{q}{\,}^\dag ,\vec{q}) \vec{q} -
(\vec{q},s_0\vec{q}) s_0\vec{q}{\,}^* =0,
\end{equation}
If we put $n=3$  and introduce the new variables $\Phi_{\pm 1}=q_{1,3}$,
$\Phi_{0}=q_2$ we recover equations (\ref{eq:1}). Likewise
with $n=5$ and $\Phi_{\pm 2}=q_{1,5}$, $\Phi_{\pm 1}=q_{2,4}$, $\Phi_{0}=q_3$
we find eq. (\ref{eq:2}).
The Hamiltonians for the MNLS equations (\ref{eq:4.2})  are given by
\begin{eqnarray}\label{eq:Ham1}
H_{{\rm MNLS}}=\int_{-\infty}^\infty d x
\left((\partial_{x}\vec{p}^T,\partial_{x}\vec{q})- (\vec{p}^T,\vec{q})^2+ \frac{1}{2}
(\vec{p}^T,s_0\vec{p})(\vec{q}^T,s_{0}\vec{q})\right),
\end{eqnarray}

\section{The Direct and the Inverse scattering problem }\label{sec:3}

\subsection{The fundamental analytic solution}

Herein we remind some basic features of the inverse scattering theory for the
operator $L$ (\ref{eq:3.2}), see  \cite{86,84}. There we have made use of the general theory
developed in \cite{Sha,ZMNP,LMP6,VSG2} and the references therein.
The Jost solutions  of $L$ are defined by:
\begin{equation}
\lim_{x \to -\infty} \phi(x,t,\lambda) e^{  i \lambda J x
}=\openone, \qquad  \lim_{x \to \infty}\psi(x,t,\lambda) e^{  i
\lambda J x } = \openone
\end{equation}
and the scattering matrix $T(\lambda,t)\equiv
\psi^{-1}\phi(x,t,\lambda)$. The special choice of $J$ and the
fact that the Jost solutions and the scattering matrix take values
in the group $SO(n+2)$ we can use the following block-matrix
structure of $T(\lambda,t)$
\begin{equation}\label{eq:25.1}
T(\lambda,t) = \left( \begin{array}{ccc} m_1^+ & -\vec{b}^-{}^T & c_1^- \\
\vec{b}^+ & {\bf T}_{22} & - s_0\vec{B}^- \\ c_1^+ & \vec{B}^+{}^Ts_0 & m_1^- \\
\end{array}\right), \qquad \hat{T}(\lambda,t) = \left( \begin{array}{ccc} m_1^- & \vec{B}^-{}^T & c_1^- \\
-\vec{B}^+ & {\bf \hat{T}}_{22} &  s_0\vec{b}^- \\ c_1^+ & -\vec{b}^+{}^Ts_0 & m_1^+ \\
\end{array}\right),
\end{equation}
where $\vec{b}^\pm (\lambda,t)$ and $\vec{B}^\pm (\lambda,t)$ are
$n$-component vectors, ${\bf T}_{22}(\lambda)$ is $n \times n$ block matrix, and $m_1^\pm
(\lambda)$, and $c_1^\pm (\lambda)$ are scalar functions. Such
parametrization is compatible with the generalized Gauss
decompositions of $T(\lambda)$ which read  as follows:
\begin{equation*}\label{eq:25.1'}
\begin{aligned}
T(\lambda,t) &= T^-_J D^+_J \hat{S}^+_J , &\quad  T(\lambda,t) &= T^+_J D^-_J \hat{S}^-_J ,\\
T^\mp_J &= e^{\pm \left(\vec{\rho}^\pm , \vec{E}^\mp_1 \right)}  , &
S^\pm _J &= e^{\pm \left(\vec{\tau}^\pm , \vec{E}^\pm_1 \right)} ,  &
D_J^\pm = \diag \left( (m_1^\pm)^{\pm 1} , {\bf m}_2^\pm , (m_1^\pm)^{\mp1} \right).
\end{aligned}
\end{equation*}
The functions $m_1^\pm$ and $n\times n$ matrix-valued) functions ${\bf m}_2^\pm$ are
are analytic for $\lambda\in\bbbc_\pm$. We have introduced also the 
notations:
\begin{equation*}\label{eq:25.1a}
\begin{aligned}
\vec{\rho}^- &=\frac{\vec{B}^-}{m_1^-}, \qquad \qquad \vec{\tau}^-
=\frac{\vec{B}^+}{m_1^-}, & \vec{\rho}^+
&=\frac{\vec{b}^+}{m_1^+}, \qquad \qquad \vec{\tau}^+
=\frac{\vec{b}^-}{m_1^+},\\
c_1^\pm &= \frac{m_1^\pm (\vec{\rho}^{\pm T} s_0 \vec{\rho}^\pm)}{2}  =
\frac{m_1^\mp (\vec{\tau}^{\mp T} s_0 \vec{\tau}^\mp)}{2} , & \\
\vec{b}^- &= \frac{\mu_2^{-,T}}{m_1^-} \vec{B}^-, \qquad \vec{b}^+
= \frac{\mu_2^{-}}{m_1^-} \vec{B}^+, &  \vec{B}^+ &=
\frac{s_0\mu_2^{+,T}s_0}{m_1^+} \vec{b}^+, \qquad \vec{B}^- =
\frac{s_0\mu_2^{+}s_0}{m_1^+} \vec{b}^-,
\end{aligned}
\end{equation*}
where $\mu^+ =\m_2^+ -\vec{b}^+ \vec{b}^{-,T}/(2m_1^{+})$,
$\mu^- =\m_2^- -s_0\vec{B}^+ \vec{B}^{-,T}s_0/(2m_1^{-})$.
There are  some additional relations which ensure that
both $T(\lambda)$ and its inverse $\hat{T}(\lambda)$ belong to the
orthogonal group $SO(n+2)$ and that $T(\lambda)\hat{T}(\lambda) =\openone$.

Important tools for reducing the ISP to a Riemann-Hilbert problem
(RHP) are the fundamental analytic solution (FAS) $\chi^{\pm}
(x,t,\lambda )$. We will introduce two pairs of FAS using  the generalized
Gauss decomposition of $T(\lambda,t)$, see \cite{ZMNP,VSG2,TMF98}:
\begin{equation}\label{eq:FAS_J}
\begin{split}
\chi ^\pm(x,t,\lambda)= \phi (x,t,\lambda) S_{J}^{\pm}(t,\lambda )
= \psi (x,t,\lambda ) T_{J}^{\mp}(t,\lambda ) D_J^\pm (\lambda), \\
\chi ^{\prime,\pm}(x,t,\lambda)= \phi (x,t,\lambda) S_{J}^{\pm}(t,\lambda )
\hat{D}_J^\pm (\lambda) = \psi (x,t,\lambda ) T_{J}^{\mp}(t,\lambda ) .
\end{split}
\end{equation}
More precisely, this construction ensures that
$\xi^\pm(x,\lambda)=\chi^\pm(x,\lambda) e^{i\lambda Jx}$ and
$\xi^{\prime,\pm}(x,\lambda)=\chi^{\prime,\pm}(x,\lambda) e^{i\lambda Jx}$
are analytic functions of $\lambda$ for $\lambda \in \bbbc_\pm$.
If $Q(x,t) $ is a solution of the MNLS eq. (\ref{eq:4.2}) then  the
matrix  elements of $T(\lambda)$  satisfy the  linear evolution equations \cite{86,84}
\begin{equation}\label{eq:evol}
\begin{aligned}
i\frac{d\vec{b}^{\pm}}{d t} \pm \lambda ^2
\vec{b}^{\pm}(t,\lambda ) &=0, & \qquad i\frac{d\vec{B}^{\pm}}{d t}
\pm \lambda ^2 \vec{B}^{\pm}(t,\lambda ) &=0, \\
i\frac{d m_1^{\pm}}{d t}  &=0, &\qquad  i \frac{d{\bf
m}_2^{\pm}}{d t}  &=0.
\end{aligned}
\end{equation}
Thus the block-diagonal matrices $D^{\pm}(\lambda)$ can be
considered as generating functionals of the integrals of motion.
The fact that all $(2r-1)^2$ matrix elements of
$\m_2^\pm(\lambda)$ for $\lambda \in \bbbc_\pm$  generate
integrals of motion reflect the superintegrability of the model
and are due to the degeneracy of the dispersion law of
(\ref{eq:4.2}). We remind that $D^\pm_J(\lambda)$ allow analytic
extension for $\lambda\in \bbbc_\pm$ and that their zeroes and
poles determine the discrete eigenvalues of $L$.

\subsection{The Riemann-Hilbert Problem}
The FAS for real $\lambda$ are linearly related \cite{86,84}
\begin{equation}\label{eq:rhp0}
\begin{aligned}
\chi^+(x,t,\lambda) &=\chi^-(x,t,\lambda) G_{0,J}(\lambda,t), &
\qquad G_{0,J}(\lambda,t)&=\hat{S}^-_J(\lambda,t)S^+_J(\lambda,t) \\
\chi^{\prime,+}(x,t,\lambda) &=\chi^{\prime,-}(x,t,\lambda) G'_{0,J}(\lambda,t),
&\qquad G'_{0,J}(\lambda,t) &=\hat{T}^+_J(\lambda,t)T^-_J(\lambda,t) .
\end{aligned}
\end{equation}
One can rewrite eq. (\ref{eq:rhp0}) in an equivalent form for the
FAS $\xi^\pm(x,t,\lambda)=\chi^\pm (x,t,\lambda)e^{i\lambda Jx }$ and
$\xi^{\prime,\pm}(x,t,\lambda)=\chi^{\prime,\pm} (x,t,\lambda)e^{i\lambda Jx }$
which satisfy the equation:
\begin{equation}\label{eq:xi}
\begin{split}
i\frac{d\xi^\pm}{dx} + Q(x)\xi^\pm(x,\lambda) -\lambda [J,
\xi^\pm(x,\lambda)]=0, \\
i\frac{d\xi^{\prime,\pm}}{dx} + Q(x)\xi^{\prime,\pm}(x,\lambda) -\lambda [J,
\xi^{\prime,\pm}(x,\lambda)]=0
\end{split}
\end{equation}
and  the relations
\begin{equation}\label{eq:rh-n}
\lim_{\lambda \to \infty} \xi^\pm(x,t,\lambda) = \openone, \qquad
\lim_{\lambda \to \infty} \xi^{\prime,\pm}(x,t,\lambda) = \openone.
\end{equation}
Then these FAS satisfy the RHP's
\begin{equation}\label{eq:rhp1}
\begin{aligned}
\xi^+(x,t,\lambda) &=\xi^-(x,t,\lambda) G_J(x,\lambda,t), &\quad
G_{J}(x,\lambda,t) &=e^{-i\lambda J(x+\lambda t)}G^-_{J}(\lambda,t)e^{i\lambda J(x+\lambda t)} ,\\
\xi^{\prime,+}(x,t,\lambda) &=\xi^{\prime,-}(x,t,\lambda) G_J(x,\lambda,t), &\quad
G'_{J}(x,\lambda,t) &=e^{-i\lambda J(x+\lambda t)}G'_{J}(\lambda,t)e^{i\lambda J(x+\lambda t)} .
\end{aligned}
\end{equation}
Obviously the sewing function $G_J(x,\lambda,t)$ (resp. $G'_J(x,\lambda,t)$) is uniquely
determined by the Gauss factors $S_J^\pm (\lambda,t)$ (resp. $T_J^\pm (\lambda,t)$). In
addition Zakharov-Shabat's theorem \cite{ZaSh} states that if
sewing functions $G_J(x,\lambda,t)$ and $G_J'(x,\lambda,t)$ depend
on $x$ and $t$ in the way prescribed above ensures that the corresponding FAS satisfy the
linear systems (\ref{eq:xi}).

Assume we have solved the RHP's above and know the FAS $\xi^+(x,t,\lambda)$.
Then the corresponding potential of $L$ is recovered by
\begin{equation}\label{eq:XI-Q}
    Q(x,t) = \lim_{\lambda\to\infty} \lambda \left( J- \xi^+(x,t,\lambda)J\hat{\xi}^+(x,t,\lambda)
    \right).
\end{equation}

\section{Reductions of MNLS}

The reduction group proposed by Mikhailov \cite{Mikh} provides four classes of reductions
which are automatically compatible with the Lax representation of the corresponding MNLS eq.

The reduction group $G_R $ is a finite group which preserves the
Lax representation $[L,M]=0$, i.e. it ensures that the reduction
constraints are automatically compatible with the evolution. $G_R $ must
have two realizations: i) $G_R \subset {\rm Aut}\fr{g} $ and ii) $G_R
\subset {\rm Conf}\, \Bbb C $, i.e. as conformal mappings of the complex
$\lambda $-plane. To each $g_k\in G_R $ we relate a reduction
condition for the Lax pair as follows \cite{Mikh}:
\begin{equation}\label{eq:2.1}
C_k(L(\Gamma _k(\lambda ))) = \eta _k L(\lambda ), \quad
C_k(M(\Gamma _k(\lambda ))) = \eta _k M(\lambda ),
\end{equation}
where $C_k\in \mbox{Aut}\; \fr{g} $ and $\Gamma _k(\lambda )\in
\mbox{Conf\,} \bbbc $ are the images of $g_k $ and $\eta _k =1 $ or $-1 $
depending on the choice of $C_k $. Since $G_R $ is a finite group then for
each $g_k $ there exist an integer $N_k $ such that $g_k^{N_k} =\openone
$. In all the cases below $ N_k=2 $ and the reduction group is isomorphic
to $\bbbz_2 $. More specifically the automorphisms $C_k $, $k=1,\dots,4 $ listed above
lead to the four possible classes of reductions for the matrix-valued functions
\begin{equation}\label{eq:U-V}
U(x,t,\lambda ) = Q(x,t) - \lambda J, \qquad
V(x,t,\lambda ) = V_0(x,t) + \lambda V_1(x,t) -\lambda^2J,
\end{equation}
of the Lax representation:
\begin{equation}\label{eq:U-V.a}
\begin{aligned}
&\mbox{1)} &\qquad C_1(U^{\dagger}(\kappa _1(\lambda )))&= U(\lambda ),
&\qquad C_1(V^{\dagger}(\kappa _1(\lambda )))&= V(\lambda ), \\
&\mbox{2)}  &\qquad C_2(U^{T}(\kappa _2(\lambda ))) &= -U(\lambda ), &\qquad
C_2(V^{T}(\kappa _2(\lambda ))) &= -V(\lambda ), \\
&\mbox{3)} &\qquad C_3(U^{*}(\kappa _1(\lambda ))) &= -U(\lambda ), &\qquad
C_3(V^{*}(\kappa _1(\lambda ))) &= -V(\lambda ), \\
&\mbox{4)} &\qquad C_4(U(\kappa _2(\lambda ))) &= U(\lambda ), &\qquad
C_4(V(\kappa _2(\lambda ))) &= V(\lambda ),
\end{aligned}
\end{equation}
In what follows  we will examine the typical reductions of MNLS eqs. of the class 1)
obtained by specifying $\kappa_1(\lambda)=\lambda^*$ and
$C_1$ to be a $\bbbz_2$-automorphism of $\fr{g}$ such that $C_1(J)=J$. Below
we list several choices for $C_1$ leading to inequivalent reductions:
\begin{equation}\label{eq:C1-1}
\begin{aligned}
\mbox{a)} \quad C_1 &=\openone, &\quad \vec{p}(x) &=\vec{q}{\,}^*(x) , \quad
&\mbox{b)} \quad C_1&=K_1, & \vec{p}(x) &=K_{01}\vec{q}{\,}^*(x) , \\
\mbox{c)} \quad C_1 &=S_{e_2}, & \quad \vec{p}(x) &=K_{02}\vec{q}{\,}^*(x) , \quad &\mbox{d)}
\quad C_1&=S_{e_2}S_{e_3},  & \vec{p}(x) &=K_{03}\vec{q}{\,}^*(x) ,
\end{aligned}
\end{equation}
where
\begin{equation}\label{eq:C1-2}
K_j =\bdiag(1, K_{0j}, 1), \qquad K_{01} = \diag(\epsilon_1,\dots ,\epsilon_{r-1},1, \epsilon_{r-1}, \dots,
\epsilon_1 ),
\end{equation}
and $\epsilon_j=\pm 1$.
The matrices $K_{02}$ and $K_{03}$ corresponding to the Weyl reflections $S_{e_2}$, $S_{e_2}S_{e_3}$ etc.
are not diagonal; they have dimension $n\times n$ and for $n=3,4$ and $5$   are given by:
\begin{equation}\label{eq:C1-3}
\begin{aligned}
n&=3, & K_{02} &= \left( \begin{array}{ccc} 0 & 0 & 1 \\ 0 & -1 & 0 \\ 1 & 0 & 0 \end{array}\right), & \\
n &=4, & K_{02}& = \left( \begin{array}{cccc} 0 & 0  & 0 & -1 \\ 0 & 1  & 0 & 0 \\
0 & 0  & 1 & 0  \\ -1 & 0  & 0 & 0  \end{array}\right), &
K_{03} &= \left( \begin{array}{cccc} 0 & 0  & 0 & 1 \\ 0 & 0  & -1 & 0 \\
0 & -1 & 0 & 0  \\ 1 & 0 &  0 & 0  \end{array}\right),\\
n &=5, & K_{02} &= \left( \begin{array}{ccccc} 0 & 0 & 0 & 0 & -1 \\ 0 & 1 &0 & 0 & 0 \\
0 & 0 & -1 & 0 & 0 \\ 0 & 0 & 0 & 1 & 0  \\ -1 & 0 & 0 & 0 & 0  \end{array}\right), &
K_{03} &= \left( \begin{array}{ccccc} 0 & 0 & 0 & 0 & -1 \\ 0 & 0 &0 & 1 & 0 \\
0 & 0 & 1 & 0 & 0 \\ 0 & 1 & 0 & 0 & 0  \\ -1 & 0 & 0 & 0 & 0  \end{array}\right),
\end{aligned}
\end{equation}

Each of the above reductions impose constraints on the FAS, on the scattering matrix $T(\lambda)$ and on its
Gauss factors $S^\pm_J(\lambda)$, $T^\pm_J(\lambda)$ and $D^\pm_J(\lambda)$. These have the form:
\begin{equation}\label{eq:C1-3'} 
\begin{aligned}
(\phi(x,\lambda^*))^\dag &= K_j^{-1}\hat{\phi}(x,\lambda) K_j,  &\quad (\psi(x,\lambda^*))^\dag &=
K_j^{-1}\hat{\psi}(x,\lambda)K_j,   \\
(\chi^+(x,\lambda^*))^\dag &=K_j^{-1}\hat{\chi}^-(x,\lambda)K_j  &\quad (T(\lambda^*))^\dag &= K_j^{-1}\hat{T}(\lambda)K_j, \\
(S^+(\lambda^*))^\dag &= K_j^{-1}\hat{S}^-(\lambda)K_j  &\quad (T^+(\lambda^*))^\dag &=K_j^{-1}\hat{T}^-(\lambda)K_j  \\
(D^+(\lambda^*))^\dag &= K_j^{-1}\hat{D}^-(\lambda)K_j  &
\end{aligned}
\end{equation}
where the matrices $K_j$ are specific for each choice of the automorphism $C_1$, see eq. (\ref{eq:C1-2}).
In particular, from the last line of (\ref{eq:C1-3'}) and (\ref{eq:C1-2}) we get:
\begin{equation}\label{eq:m1pm}
(m_1^+(\lambda^*))^* = m_1^-(\lambda),
\end{equation}
and consequently, if $m_1^+(\lambda)$ has zeroes at the points $\lambda_k^+$, then
$m_1^-(\lambda)$ has  zeroes at:
\begin{equation}\label{eq:lapm}
\lambda_k^- = (\lambda_k^+)^*, \qquad k=1,\dots, N.
\end{equation}

Below we will write down the effects of these reductions on the corresponding Hamiltonians.
For the typical reduction $\vec{p}= \vec{q}{\,}^*$ we get:
\begin{eqnarray}\label{eq:Ham2}
H_{{\rm MNLS}}=\int_{-\infty}^\infty d x \left\{
\left((\partial_{x}\vec{q},\partial_{x}\vec{q}{\,}^{*}\right)-
(\vec{q},\vec{q{\,}^{*}})^2+ \frac{1}{2} (\vec{q},s_0\vec{q})(\vec{q^{*}},s_{0}\vec{q^{*}})\right\},
\end{eqnarray}

\begin{eqnarray}\label{eq:Ham1'}
H_{{\rm MNLS}}^{(j)}=\int_{-\infty}^\infty d x \left((\partial_{x}\vec{q}K_j\partial_{x}\vec{q^{*}})-
(\vec{q},K_j\vec{q^{*}})^2+ \frac{1}{2}(\vec{q},s_0\vec{q})(\vec{q^{*}},s_{0}\vec{q^{*}})\right),
\end{eqnarray}
The Hamiltonian $H_{{\rm MNLS}}^{(1)}$ with $K_{01}$ (\ref{eq:C1-2}) has indefinite kinetic term. As
a consequence the corresponding MNLS has singular soliton solutions which `blow-up' in
finite time.

The above Hamiltonians, after the change of variables can be written in more `aesthetic' form. Indeed, for
odd $n=2r-1$ we can put:
\begin{equation}\label{eq:q-v}
q_{2k-1, 2r-2k+1} = \frac{v_{2k-1} \pm i v_{2r-2k+1}}{\sqrt{2}}, \qquad
q_{2k, 2r-2k} = \frac{i v_{2k} \mp v_{2r-2k}}{\sqrt{2}}, \qquad  q_r= c_{0,r}v_r;
\end{equation}
with $k=1,2,\dots, r-1$ and $c_{0,r}=e^{(r-1)\pi i/2}$; for $n=2r$ we put:
\begin{equation}\label{eq:q-v'}
q_{2k-1, 2r-2k+2} = \frac{v_{2k-1} \pm i v_{2r-2k+2}}{\sqrt{2}}, \qquad
q_{2k, 2r-2k+1} = \frac{iv_{2k} \mp  v_{2r-2k+1}}{\sqrt{2}}
\end{equation}
with $ k=1,2,\dots, r$.

Inserting the above changes of variables into the Hamiltonian (\ref{eq:Ham2}) we get
\begin{eqnarray}\label{eq:Ham3}
H_{{\rm KS}}= \int_{-\infty}^\infty d x \left\{ \sum_{j=1}^n|\partial_{x} v_j|^2  -
\left(\sum_{j=1}^n |v_j|^2\right)^2 + \frac{1}{2} \left| \sum_{j=1}^n v_j^2\right|^2  \right\},
\end{eqnarray}
which is the Hamiltonian of the $n$-component Kulish-Sklyanin model (KSM) \cite{KuSkl}. Thus we have demonstrated that the Lax pairs
(\ref{eq:3.1}), (\ref{eq:3.2}) can be used  also for integrating the MNLS (\ref{eq:Ham3}).
In their original paper \cite{KuSkl} Kulish and Sklyanin have used Lax pair whose potential is an element
of a Clifford algebra. Later Sokolov and Svinolupov \cite{SviSok} discovered another class of Lax pairs for these  models whose
potentials take values in Jordan algebras. The above Lax pairs allowed to prove integrability of
the KSM but were not convenient for solving the inverse scattering problem and constructing exact solutions.
Another important property of these models is that
they possess both classical \cite{ForKu*83} and quantum $R$-matrices \cite{KuSkl}.

Another way to obtain KSM is to apply the reduction of type 4) with $K_0=\bdiag (1, \epsilon s_0, 1)$,
where $\epsilon =\pm 1$. For odd values of $n=2r-1$ this reduction means that:
\begin{equation}\label{eq:qs0q}
q_k = (-1)^{k+1} \epsilon q_{2r-k} =w_k, \qquad k=1,\dots , r,
\end{equation}
while for $n=2r$ one gets:
\begin{equation}\label{eq:qs0qe}
q_k = (-1)^{k+1} \epsilon q_{2r-k+1} =w_k, \qquad k=1,\dots , r.
\end{equation}
This reduction leads to $r$-component  KSM.

Let us write down the Hamiltonians for the different  reductions. Below for convenience we
will split $H_{{\rm MNLS}}$ into kinetic and interaction terms: $H_{{\rm MNLS}} =H_{{\rm kin}}^{(j)}
-H_{{\rm int}}^{(j)}$.

Reduction b):
\begin{equation}\label{eq:H1}
\begin{split}
H_{{\rm kin}}^{(1)} &= \int_{-\infty}^\infty d x \left\{ \sum_{j=1}^{r-1} \epsilon_j (|\partial_{x} q_j|^2 +
|\partial_{x} q_{2r-j}|^2) + |\partial_{x} q_r|^2\right\}, \\
H_{{\rm int}}^{(1)} &=  \int_{-\infty}^\infty d x \left\{ \sum_{j=1}^{r-1} \epsilon_j (|q_j|^2 + |q_{2r-j}|^2)
+ |q_r|^2)^2  \right.
\\  & \qquad \qquad \left.  -\frac{1}{2} \left| \sum_{j=1}^{r-1} (-1)^{j+1} 2q_jq_{2r-j} + (-1)^r q_r^2\right|^2   \right\} ,
\end{split}
\end{equation}

One can construct other  reductions, e.g. ones  of type c) with reduction matrix $K_j$. Then
\begin{equation}\label{eq:H2c}
H_{{\rm MNLS}}^{(j)} = \int_{-\infty}^\infty d x \left\{  ( \partial_{x} \vec{q}{\,}^\dag K_j \partial_{x} \vec{q})
- (\vec{q}{\,}^\dag K_j \vec{q})^2  +\frac{1}{2} \left| (\vec{q}^T s_0 \vec{q})  \right|^2   \right\} ,
\end{equation}

Characteristic feature of the reductions involving Weyl group elements is that they lead to `non-diagonal'
form of the kinetic terms \cite{GKMV*02}. Making simple change of variables  diagonalizing $K_j$ we can recover the
diagonal form of the kinetic terms but unfortunately we can not make it positive definite. This is related to the
fact that $K_j^2=1$ and so has as eigenvalues both $+1$ and $-1$ with certain multiplicities.

Let us give also an important example of class 2) reductions (\ref{eq:U-V}).
The constraints that these class of reductions impose on the FAS and on the scattering matrix $T(\lambda)$ and on its
Gauss factors $S^\pm_J(\lambda)$, $T^\pm_J(\lambda)$ and $D^\pm_J(\lambda)$ take the form:
\begin{equation}\label{eq:C1-3''}
\begin{aligned}
(\chi^+(x,\lambda))^T &=K_j^{\prime,-1}\hat{\chi}^-(x,\lambda)K'_j  &\quad (T(\lambda))^T &= K_j^{\prime,-1}\hat{T}(\lambda)K'_j, \\
(S^\pm(\lambda))^T &= K_j^{\prime,-1}\hat{S}^\pm(\lambda)K'_j  &\quad (T^\pm(\lambda))^T &=K_j^{\prime,-1}\hat{T}^\pm (\lambda)K'_j  \\
 \end{aligned}
\end{equation}
and $(D^\pm(\lambda))^T = K_j^{\prime,-1}\hat{D}^\pm(\lambda)K'_j  $.
The explicit form of the matrices $K_j'$ is determined by the particular realization of the automorphism $C_2$.
Choosing $n=3$ and $C_2=S_{e_1}$ we obtain the constraint $q_1=q_3$ and the  reduced Hamiltonian takes the form:
\begin{equation}\label{eq:HH}
H_{{\rm }}= \int_{-\infty}^\infty d x \left\{ |\partial_{x} v_1|^2 + |\partial_{x} v_2|^2  -
( |v_1|^2 + |v_2|^2)^2 +\frac{1}{2} \left| v_1^2 - v_2^2 \right|^2  \right\},
\end{equation}
where we have put $q_1=q_2 = \frac{1}{\sqrt{2}} v_1$ and $q_2=v_2$. This model also been derived as
relevant for $F=1$ BEC \cite{Kevre}.

\section{Dressing method and soliton solutions}
The  dressing Zakharov-Shabat method  \cite{ZaSh,ZaMi} for constructing soliton solutions
of MNLS has been modified in \cite{84} for the BD.I-type symmetric spaces. There we also
analyzed the different types of soliton solutions.
Below we briefly discuss the properties of the generic one-soliton solutions

It is obtained by dressing the regular FAS  $\chi^\pm_0(x,\lambda)$  of the RHP (\ref{eq:rhp1}).
Using them we construct the singular solutions $\chi^\pm_0(x,\lambda)$ of the RHP
\begin{equation}\label{eq:dres}
\begin{split}
\chi^\pm(x,\lambda) &= u(x,\lambda) \chi^\pm_0(x,\lambda)\hat{u}_-, \qquad
\chi^{\prime,\pm} (x,\lambda) = u(x,\lambda) \chi^\pm_0(x,\lambda)\hat{u}_+,\\
u(x,\lambda) &= \openone + (c_1(\lambda)-1)P_1(x) + (c_1^{-1}(\lambda)-1) \bar{P}_1(x), \qquad
u_\pm=\lim_{x\to\pm\infty} u(x,\lambda).
\end{split}
\end{equation}
For the above choice of $J$ it is enough to consider rank 1 projectors $P_1(x,t)$ and $\bar{P}_1(x,t) = S_0P_1^T S_0$.
Together with the constraint $P_1\bar{P}_1 =0$, the last condition ensures that $u(x,t)\in SO(n+2)$.
It remains to only to give the explicit form of $P_1(x,t)$. Generically it is determined by two
polarization vectors  $|n_{0,1}\rangle $ and $\langle m_{0,1}|$, and the initial regular solutions:
\begin{equation}\label{eq:n-m}
\begin{aligned}
P_1(x,t) &= \frac{|n_1(x,t)\rangle \langle m_1(x,t)|}{\langle m_1(x,t)|n_1(x,t)\rangle}, &\qquad
\bar{P}_1(x,t) &= \frac{|\overline{m_1(x,t)}\rangle \langle\overline{n_1(x,t)}|}
{\langle\overline{n_1(x,t)}|\overline{m_1(x,t)}\rangle},
\end{aligned}
\end{equation}
\begin{equation}\label{eq:xxx}
\begin{aligned}
|n_1(x,t)\rangle &= \chi^+_0(x,t,\lambda_1^+) |n_{0,1}\rangle , &\qquad \langle m_1(x,t)| &=
\langle m_{0,1}| \hat{\chi}^-_0(x,t,\lambda_1^-), \\
|\overline{m_1(x,t)}\rangle &= \chi^-_0(x,t,\lambda_1^-) |\overline{m_{0,1}}\rangle , &\qquad \langle\overline{m_1(x,t)}| &=
\langle\overline{n_{0,1}}| \hat{\chi}^-_0(x,t,\lambda_1^-).
\end{aligned}
\end{equation}
The one soliton solution is parametrized by
the two eigenvalues $\lambda_1^\pm$ and by the polarization vectors  $|n_{0,1}\rangle $ and
$\langle m_{0,1}|$. The latter after renormalization have $n-1$ independent components each:
\[ |n_{0,1}\rangle  = \left( \begin{array}{c} \sqrt{A_0}, \\ \vec{\nu}_{0,1}/\sqrt{A_0} \\ 1/\sqrt{A_0} \end{array} \right),
\qquad \langle m_{0,1}| = \left(\sqrt{B_0}, \vec{\mu}_{0,1}/\sqrt{B_0}, 1/\sqrt{B_0} \right),
\]
where $A_0=\frac{1}{2} (\vec{\nu}_{0,1}^T s_0 \vec{\nu}_{0,1})$ and $B_0=\frac{1}{2} (\vec{\mu}_{0,1}^T s_0 \vec{\mu}_{0,1})$.
The constraint $P_1\bar{P}_1 =0$ means that the vectors $\vec{\mu}_{0,1}$ and $\vec{\nu}_{0,1}$ must satisfy
$\vec{\mu}_{0,1}^T s_0 \vec{\nu}_{0,1} =0$.
Therefore the one-soliton solution can be viewed as a dynamical system with $2n-1$ degrees of freedom.
After some simplifications it takes the form:
\begin{equation}\label{eq:1s}
q_k(x,t) = -\frac{4i\nu_1}{\Delta} e^{-i\mu_1 \tilde{z}_k} e^{\tilde{\xi}_{0,k}} \left[
\cos(\delta_{0k}) \cosh (z_{0k}) + i\sin(\delta_{0k}) \sinh (z_{0k}) \right] ,
\end{equation}
\begin{equation}\label{eq:1sa}
\begin{aligned}
\Delta &= 2 \cosh (2z_0) + \mathcal{C}, &\quad \mathcal{C} &= \frac{(\vec{\nu}_0{\,}^\dag \vec{\nu}_0)} {|A_0|},
&\qquad \tilde{z}_k &= x+w_1t -\tilde{\delta}_{0,k}/\mu_1, \\
z_0 &= \nu_1 (x-u_1t) +\xi_0, & z_{0k} &= \nu_1x +\xi_{0,k}, &\quad  \tilde{z}_{0k} &= \nu_1(x-v_1t) +\tilde{\xi}_{0,k}, \\
\xi_0 &= \frac{1}{2} \ln |A_0|, & \xi_{0,k} &= \frac{1}{2} \ln \frac{|\vec{\nu}_{0,2r-k}|}{|A_0||\vec{\nu}_{0,k}|},
&\quad \tilde{\xi}_{0,k} &= \frac{1}{2} \ln \frac{|\vec{\nu}_{0,k}||\vec{\nu}_{0,2r-k}|}{|A_0|},
\end{aligned}
\end{equation}
\[
\delta_{0,k} = (\alpha_{0,2r-k} + \alpha_{0,k} -\alpha_0 -\pi k)/2, \qquad
\tilde{\delta}_{0,k} = (\alpha_{0,2r-k} - \alpha_{0,k} +\alpha_0 +\pi k)/2, \]
where $\alpha_0=\arg A_0$ and $\alpha_{0,k}=\arg \vec{\nu}_{0,k}$.

Each of the reductions of the type (\ref{eq:U-V.a}) imposes constraints not only on $\lambda_1^+ =
(\lambda_1^-)^* $, but also on the polarization vectors:
\begin{equation}\label{eq:mu01}
 \vec{\mu}_{0,1} = K_{0,j} \vec{\nu}_{0,1}{\,}^* , \qquad \vec{\nu}_{0,1}^T K_{0,j}s_0 \vec{\nu}_{0,1} =0.
\end{equation}
As a result, after the reduction the number of independent parameters of the soliton solution
becomes $n-1$. The velocities $u_1$ and $w_1$ are given by $u_1= -2\mu_1$ and $w_1=(\nu_1^2-\mu_1^2)/\mu_1$.

Special attention deserves the fact that generically all $z_{0,k}$ are different and as a result each component
$q_k(x,t)$ has its center of mass shifted with respect to the others.

Let us now consider a $\bbbz_2\times \bbbz_2$ reduction by applying simultaneously two reduction: the
first is the typical one and the second is the class 2) reductions as for the model (\ref{eq:HH}). The first reduction
imposes the relation (\ref{eq:mu01}) between the two polarization vectors  $|n_{0,1}\rangle$ and $\langle n_{0,1}|$.
The second reduction imposes constraint on the vector $|n_{0,1}\rangle $, namely:
\[ |n_{0,1}\rangle = K_{0j}' |n_{0,1}\rangle .\]
In particular, for $n=3$ and $C_2=S_{e_1}$ the vector $|n_{0,1}\rangle $ has 3 components and $K_{0j}'=\diag (1,-1,1)$.
Thus only two independent complex coefficients are enough to parametrize the corresponding polarization vector,
and the corresponding soliton can be viewed as dynamical system with three degrees of freedom.

\section{Discussion and conclusions}

One of the important consequences of the FAS is that with their help
one can construct the kernel of the resolvent of $L$ (see \cite{LMP6,Holyoke})
and prove the completeness relation for its eigenfunctions. From these expressions
it becomes obvious that the resolvent develops poles at all points $\lambda_k^\pm \in \bbbc_\pm$
for which  $m_1^\pm(\lambda_k^\pm)=0$. Combining this fact with the equivalence between the
solutions of the RHP and the FAS of the Lax operator we conclude that the singularities of the
RHP correspond to the discrete eigenvalues of $L$.

Quite often  the general analysis of the MNLS (\ref{eq:1}) is followed by simplifications which
often reduce the MNLS to a single-component NLS. One way do to this was mentioned above: it is to impose the
reduction $\Phi_{+1}=\Phi_{-1}$. Another less obvious way to this is to impose this reduction on
the initial conditions. Indeed, one can show that imposing $\Phi_{+1}(x,t=0)=\Phi_{-1}(x,t=0)$ ensures
that $\Phi_{+1}(x,t)=\Phi_{-1}(x,t)$ for all $t>0$. At the same time there is a substantial difference between
the solitons of the scalar NLS or Manakov model and the solitons of MNLS (\ref{eq:1}). Unlike the solitons
of the Manakov model, all three components of the one-soliton solution of (\ref{eq:1})  have
different $x$-dependence; generically each component has different `center of mass' position. Therefore,
if one wants to demonstrate new nontrivial aspects of soliton dynamics one should use generic
initial values for $\Phi_{\pm 1}(x,t=0)$and $\Phi_{-1}(x,t=0)$.

Another still open problem is the interrelation between the solutions of the direct and inverse
scattering problem for $L$, considered in different irreducible representations (IRREP) of the corresponding
Lie algebra $\fr{g}$. From the point of view of the relevant NLEE, their Lax representations have purely
algebraic nature and therefore, the form of the NLEE {\em does not depend} on the choice of the IRREP of $\fr{g}$.

From the  point of view of the spectral theory, the different IRREP have different dimensions; therefore
changing the IRREP we change {\em the order} of the corresponding operator. Since we are dealing with simple
Lie algebras whose IRREP are well known \cite{Helg}. In particular, it is well known that the finite dimensional
representations can be realized as invariant subspaces of the tensor products of the typical one. Let us assume that we
are able to construct the FAS and the relevant RHP and dressing factors in the typical representation. Obviously,
 taking the tensors products of the FAS their analyticity properties will persist and we will get the corresponding
 FAS and RHP in the corresponding IRREP. However nontrivial things may take place when one considers the
 multiplicities of the corresponding discrete eigenvalues.

As an example I will just mention that the dressing factor  can be evaluated also for the other fundamental
representations of $\fr{g}$ \cite{PLA126}.
If in the typical representation of $\fr{g}\simeq so(2r)$ $u(x,\lambda)$ is given by (\ref{eq:dres}) then in the
spinor representation it will take the form \cite{Rosen}:
\[ u(x,\lambda) = \sqrt{c_1(\lambda)} \pi_1(x,t) + \frac{1}{\sqrt{c_1(\lambda)}} \bar{\pi}_1(x,t), \qquad
 \bar{\pi}_1(x,t) = \tilde{s}_0  \pi_1(x,t)\tilde{s}_0^{-1} , \]
and the projectors satisfy $ \pi_1(x,t)  \bar{\pi}_1(x,t)=0$ and  $ \pi_1(x,t)  +\bar{\pi}_1(x,t)=\openone$. Note the
substantial change in the $\lambda$-dependence of $u(x,\lambda)$, as well as the fact that now instead of having
rank one projectors $P_1(x,t)$ we get projectors $ \pi_1(x,t)$ and  $ \bar{\pi}_1(x,t)$ of rank $r$.

We will discuss these problems in more details elsewhere.


\begin{theacknowledgments}
It is my pleasure to thank the organizers of the AMITANS conference for kind invitation.
I am grateful to professor N. Kostov  and to an anonymous referee for useful suggestions.
\end{theacknowledgments}



\bibliographystyle{aipproc}   

\bibliography{sample}

\begin{thebibliography}{9}

\bibitem{IMW04} J. Ieda, T. Miyakawa, and M. Wadati. Exact analysis of soliton
dymamics in spinor Bose-Einstein condesates.
\emph{Phys. Rev Lett.} {\bf 93}, 194102 (2004).

\bibitem{86} V. S. Gerdjikov, N. A. Kostov, T. I. Valchev.
Solutions of multi-component NLS models and Spinor Bose-Einstein
condensates \emph{Physica D} {\bf 238}, 1306-1310 (2009); {\bf ArXiv:0802.4398 [nlin.SI]}.

\bibitem{84} V. S. Gerdjikov, D. J. Kaup, N. A. Kostov, T. I. Valchev.
{\it On classification of soliton solutions of multicomponent
nonlinear evolution equations.} \emph{J. Phys. A: Math. Theor.} {\bf
41}  315213 (2008) (36pp).

\bibitem{ForKu*83} A. P. Fordy, and P. P. Kulish.
Nonlinear {S}chr{\"o}dinger equations and simple {L}ie algebras.
\emph{Commun.\ Math.\ Phys.} {\bf 89}, 427--443 (1983).

\bibitem{Helg} Helgasson S. {\it Differential geometry, Lie groups and
symmetric spaces}, Academic Press, (1978).

\bibitem{KuSkl} P. P. Kulish, E. K. Sklyanin. $O(N)$-invariant nonlinear Schrodinger
equation - a new completely integrable system. \emph{Phys. Lett.} {\bf 84A}, 349-352 (1981).

\bibitem{Mikh} Mikhailov A V. The Reduction Problem and the Inverse
Scattering Problem. \emph{Physica D}, {\bf 3D}, no. 1/2, 73--117 (1981).

\bibitem{ZaSh} V. E. Zakharov, A. B. Shabat.
A scheme for integrating the nonlinear equations of mathematical
  physics by the method of the inverse scattering problem. I.
\emph{Functional Analysis and Its Applications}, {\bf 8}, 226--235 (1974). \\
V. E. Zakharov, A. B. Shabat. Integration of nonlinear equations of mathematical physics by the
  method of inverse scattering. II.
\emph{Functional Analysis and Its Applications}, {\bf 13}, 166--174 (1979).

\bibitem{ZaMi}
 V. E. Zakharov,  and A. V. Mikhailov. On The Integrability of Classical
Spinor Models in Two-dimensional Space-time {\it Comm. Math. Phys.}
{\bf 74},  21--40 (1980).

\bibitem{82} Nikolay Kostov, Vladimir Gerdjikov. Reductions of
multicomponent mKdV equations on symmetric spaces of {\bf DIII}-type.
\emph{SIGMA} {\bf 4} (2008), paper 029, 30 pages; {\bf ArXiv:0803.1651}.

\bibitem{ZMNP} Zakharov V E., Manakov S V., Novikov S P., Pitaevskii L I.
{\it Theory of solitons. The inverse scattering method}, Plenum, N.Y.
(1984).

\bibitem{LMP6} V.~S.~Gerdjikov.
On the spectral theory of the  integro--differential
operator $\Lambda$, generating  nonlinear  evolution equations.
\emph{Lett. Math. Phys.} {\bf 6,} n.~6, 315--324, (1982).

\bibitem{VSG2} V. S. Gerdjikov.  Generalised Fourier transforms for  the  soliton
     equations. Gauge covariant formulation.
\emph{Inverse Problems} {\bf 2}, no. 1, 51--74, (1986).

\bibitem{TMF98} V.~S.~Gerdjikov. The Generalized Zakharov--Shabat System and
the Soliton Perturbations.
\emph{Theor. Math. Phys.} {\bf 99}, No.~2, 292--299 (1994).

\bibitem{SviSok} S I Svinolupov. Second-order evolution equations with symmetries.
Russian Mathematical Surveys {\bf 40}, 241-242  (1985).\\
S. I. Svinolupov and V. V. Sokolov.
Vector-matrix generalizations of classical integrable equations
\emph{Theor. Math. Phys.} {\bf 100}, 214-218, (1994).

\bibitem{GKMV*02} V. Gerdjikov, A. Kyuldjiev, G. Marmo, G. Vilasi.
Complexifications and Real Forms of Hamiltonian Structures.
\emph{European J. Phys.} {\bf 29B}, 177-182 (2002).

\bibitem{Kevre} H. E. Nistazakis, D.J. Frantzeskakis, P.G. Kevrekidis, B.A. Malomed, and R. Carretero-Gonz´alez.
Bright-Dark Soliton Complexes in Spinor Bose-Einstein Condensates. \emph{Phys. Rev. A} {\bf 77}, 033612 (2008).

\bibitem{Holyoke} V. S. Gerdjikov.
Algebraic and Analytic Aspects of $N $-wave Type Equations.
{\bf nlin.SI/0206014}; \emph{Contemporary Mathematics} {\bf 301}, 35-68 (2002).

\bibitem{PLA126} V.~S.~Gerdjikov. The Zakharov--Shabat dressing method amd the
     representation theory of the semisimple Lie algebras.
     \emph{Phys. Lett. A,} {\bf 126A,} n.~3, 184--188, (1987).

\bibitem{Rosen} R. Ivanov. On the dressing method for the generalised
Zakharov–Shabat system. \emph{Nuclear Physics B} {\bf 694 [PM]} 509–524 (2004).

\bibitem{Sha} A. B. Shabat. Inverse-scattering problem for a system of differential equations.
\emph{Functional Analysis and Its Applications}, {\bf 9}, 244--247 (1975). \\
--- An inverse scattering problem.
\emph{Diff. Equations}, {\bf 15} 1299--1307 (1979).

\end{thebibliography}

\IfFileExists{\jobname.bbl}{}
 {\typeout{}
  \typeout{******************************************}
  \typeout{** Please run "bibtex \jobname" to optain}
  \typeout{** the bibliography and then re-run LaTeX}
  \typeout{** twice to fix the references!}
  \typeout{******************************************}
  \typeout{}
 }


\end{document}